\documentclass[aps,pra,preprint,superscriptaddress]{revtex4-1}

\usepackage{graphicx}
\usepackage{amsmath}
\usepackage{mathrsfs}
\usepackage{physics}

\allowdisplaybreaks

\begin{document}


\title{Enhanced molecular chiral signals via molecular coherence}


\author{Tuguldur Kh. Begzjav}
\affiliation{Institute for Quantum Science and Engineering, Department of Physics and Astronomy,\\ Texas A\&M University, College Station, TX 77843, USA}
\email{mn.tuguldur@tamu.edu}

\author{Zhedong Zhang}
\affiliation{Institute for Quantum Science and Engineering, Department of Physics and Astronomy,\\ Texas A\&M University, College Station, TX 77843, USA}

\author{Marlan O. Scully}
\affiliation{Institute for Quantum Science and Engineering, Department of Physics and Astronomy,\\ Texas A\&M University, College Station, TX 77843, USA}
\affiliation{Department of Physics, Baylor University, Waco, TX 76706, USA}
\affiliation{Department of Mechanical and Aerospace Engineering, Princeton University, Princeton, NJ 08544, USA}

\author{Girish S. Agarwal}
\affiliation{Institute for Quantum Science and Engineering, Department of Physics and Astronomy,\\ Texas A\&M University, College Station, TX 77843, USA}
\affiliation{Department of Biological and Agricultural Engineering, Texas A\&M University, College Station, TX 77843, USA}


\date{\today}

\begin{abstract}
One of the most widely used chiroptical spectroscopic methods for studying chiral molecules is Raman optical activity; however, the chiral Raman optical activity signal is extremely weak. Here, we theoretically examine enhanced chiral signals in a system with strongly prepared molecular coherence. We show that the enhanced chiral signal due to strong molecular coherence is up to four orders of magnitude higher than that of the spontaneous Raman optical activity. We discuss several advantages of studying the heterodyned signal obtained by combining the anti-Stokes signal with a local oscillator. The heterodyning allows direct measurement of the ratio of the chiral and achiral parameters. Taking advantage of the molecular coherence and heterodyne detection, the coherent anti-Stokes Raman scattering technique opens up a new potential application for investigation of biomolecular chirality.
\end{abstract}


\maketitle

\section{Introduction}
Raman scattering from chiral molecules results in different scattered intensities for right- and left-circularly polarized incident light. This is the fundamental concept of the chiroptical spectroscopic tool called Raman optical activity (ROA) \cite{Barron1971,Barron1973,Hug2003,Barron2004,Nafie2011,Parchansky2014,Barron2015review}. Since the pioneering works by L.~D.~Barron et al. in the early 1970s, ROA has been of great interest due to its potential applications in the study of biomolecules. Applications of ROA are widespread since it allows us to retrieve molecular structural and conformational information through spectral analysis of vibrational modes of chiral biomolecules that is sometimes unobtainable by other methods. Nowadays, the ROA method is advanced enough and commercialized \cite{Hug2003}. However, the intensity of the chiral signal provided by ROA is not strong enough due to weak magnetic dipole and electric quadrupole interactions. In spite of extensive studies on chirality via ROA, enhancing the chiral signal is still a challenge. Thus one is constantly examining newer methods for the study of chiral signals \cite{Koroteev1995,Nafie1996,Zheltikov1999,Merten2010}.

Coherent anti-Stokes Raman scattering (CARS) is known to be extremely useful in studying molecular vibrations and has several advantages over spontaneous Raman spectroscopy \cite{Cheng2004}. Thus, it has been argued that it is better to study chirality by using CARS \cite{Bjarnason1980,Oudar1982} and we might refer to this as CARS-ROA. The first experimental realization of infrared- as well as visible-excited CARS-ROA was reported in Refs.~\cite{Hiramatsu2012,Hiramatsu2013,Hiramatsu2015}. It was found that the contrast of the visible-excited CARS-ROA spectrum of ($-$)-$\beta$-pinene compared with spontaneous ROA measurement can be higher by two orders of magnitude \cite{Hiramatsu2015}.

\begin{figure}[]
\begin{center}
\includegraphics[scale=1]{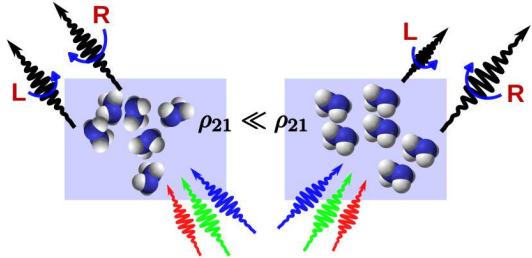}
\caption{Schematic illustration of CARS-ROA.}\label{fig_intro}
\end{center}
\end{figure}

It is well known that typical CARS signal is coherent and its magnitude is several times stronger than spontaneous Raman signal due to molecular coherence\cite{Petrov2007,Cheng2001,Volkmer2002,Pestov2008}. Consequently, it seems to be that enhancement due to molecular coherence is also valid for chiral nonlinear signals. This question is still unanswered. Thus, unlike early theoretical works \cite{Bjarnason1980,Oudar1982} on CARS-ROA which consider complete electrodynamical treatment of the problem, we focus on molecular coherence and its role on enhancement of CARS-ROA signals.
Particularly, we develop theoretical model for CARS-ROA and show how molecular coherence allow us to obtain stronger CARS-ROA signal (see Fig.~\ref{fig_intro}).
In our model, CARS-ROA process is separated into two parts in sequential time; one with preparing the molecular system with well-defined coherence and another with using a laser field to scatter from molecular coherence to produce a chiral anti-Stokes signal. It is estimated that the CARS-ROA signal is four orders of magnitude bigger than that produced by spontaneous Raman techniques. We also discuss many advantages of heterodyning the CARS-ROA signal with a local oscillator at the anti-Stokes frequency and of pre- and post-choosing the polarizations. Heterodyning gives the most direct measurement of the chirality coefficient, especially the one arising from a nonvanishing magnetic dipole contribution.

\section{Theoretical model of CARS-ROA}
\begin{figure}[]
\begin{center}
\includegraphics[scale=1]{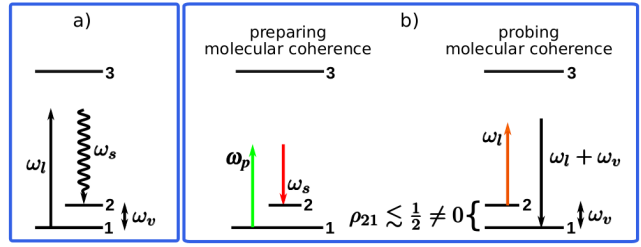}
\caption{Transition paths in the molecular level scheme for various spectrocopic techniques. a) Spontaneous ROA b) CARS-ROA.}\label{fig1}
\end{center}
\end{figure}

In this section we present our model for the enhancement of CARS-ROA signals. The generation of the spontaneous ROA signal is depicted in Fig.~\ref{fig1}a.
The system is excited by a pump beam of frequency $\omega_l$ and the scattered beam (or the spontaneously generated radiation) carries the information about the chirality of the molecule. The chiral contributions arising from the magnetic dipole and the quadrupole contributions have been extensively evaluated \cite{Barron1971}.
On the other hand, K.~Hiramatsu et al. \cite{Hiramatsu2012,Hiramatsu2013,Hiramatsu2015} reported the observation of ROA signals via coherent anti-Stokes Raman scattering (see Fig.~\ref{fig1}b). The generated signal at $2\omega_l-\omega_s$ is coherent and carries signatures of the Raman optical activity.

Guided by the advantages of molecular coherence \cite{Harris1998} we consider that the molecular system has been prepared in a coherent superposition of the levels $\vert 1\rangle$ and $\vert 2\rangle$. This can be done by using ultrashort pulses (much shorter than the transverse relaxation time $T_2$ for the molecule). This initial preparation has the advantage of preparing all the molecules vibrating in unison. Thus, the situation we consider is shown in Fig.~\ref{fig1}b where we assume the molecular system with moderate amount of molecular coherence. Note that the maximum allowed value of coherence is $1/2$. The molecular coherence has a frequency close to the vibrational frequency $\omega_v$, and it will decay as $\exp{(-t/T_2)}$. We next scatter the laser field of frequency $\omega_l$ for molecular coherence to produce a coherent scattered signal at the anti-Stokes frequency $\omega_{as}=\omega_l+\omega_v$. Advantages of using molecular coherence in CARS-ROA measurement are discussed after we present expressions for the signals. For the model of Fig.~\ref{fig1}b we need to calculate the induced polarization and magnetization to first order in the field $E_l$. It is necessary to include both dipole and quadrupole contributions. The signal at frequency $\omega_{as}$ can then be obtained using the Maxwell equations.

The semiclassical Hamiltonian of the molecule--field system is written as
\begin{align}\label{Hamiltonian0}
\hat{H}=\hat{H}_0+\hat{H}_\text{int},
\end{align}
where
\begin{align}\label{Hamiltonian}
\hat{H}_0=&\sum_r \hbar\omega_r\vert r\rangle\langle r\vert, \nonumber\\
\hat{H}_{\text{int}}=&-\hat{\mathbf{\pmb{\mu}}}\cdot \mathbf{E}_l(t)-\hat{\mathbf{m}}\cdot \mathbf{B}_l(t)
-\frac{1}{3}\sum_{\alpha,\beta}\hat{q}_{\alpha\beta}\nabla_\alpha E_{l,\beta} (t).
\end{align}
The free Hamiltonian of a molecule with transition frequency $\omega_r$ is denoted by $\hat{H}_0$ and $\{\vert r\rangle\}$ are molecular electronic and vibrational states. The interaction Hamiltonian $\hat{H}_{\text{int}}$ consists of three terms, namely electric dipole $\hat{\pmb{\mu}}$, magnetic dipole $\hat{\mathbf{m}}$ and electric quadrupole moment $\hat{q}_{\alpha\beta}$.
The electric and magnetic fields of the incident laser pulse at time $t$ are denoted by $\mathbf{E}_{l}(t)$ and $\mathbf{B}_{l}(t)$, respectively. Without loss of generality, we choose the $z$-axis as propagation direction of the incident laser pulse throughout this paper. First, let us assume the incident laser pulse to be $x$-polarized and denote it by $E_{l,x}(t)$.
We choose this electric field of incident laser pulse to have center frequency $\omega_l^{(0)}$ and delay time $\tau$ as $E_{l,x}(t)=\mathcal{E}_{l,x}(t-\tau){\rm e}^{-i\omega_l^{(0)} t}$ in the time domain and $E_{l,x}(\omega_l)=\mathcal{E}_{l,x}(\omega_l-\omega_l^{(0)}){\rm e}^{i(\omega_l-\omega_l^{(0)})\tau}$ in the frequency domain.

Using first-order perturbation theory, the $\beta$th components of induced dipole and quadrupole moments are found to be
\begin{align}\label{induced}
\mu^{(e)}_\beta(\omega_{as})&= \frac{1}{\pi} \int_{-\infty}^{\infty}\dd{\omega_l} \tilde{\alpha}_{\beta\alpha}(\omega_{as}-\omega_{v}) \frac{\mathcal{E}_{l,\alpha} (\omega_l-\omega_{l}^{(0)})
{\rm e}^{i(\omega_l-\omega_l^{(0)})\tau}}{\Gamma^2+(\omega_{as}-\omega_{v}-\omega_l)^2}
\Gamma\rho_{21}(0),\nonumber\\
\mu^{(m)}_\beta(\omega_{as})&= \frac{1}{\pi} \int_{-\infty}^{\infty}\dd{\omega_l}\tilde{G}_{\beta\alpha}(\omega_{as}-\omega_{v}) \frac{\mathcal{B}_{l,\alpha} (\omega_l-\omega_{l}^{(0)}){\rm e}^{i(\omega_l-\omega_l^{(0)})\tau}}{\Gamma^2+(\omega_{as}-\omega_{v}-\omega_l)^2}\Gamma\rho_{21}(0),\nonumber\\
m_\beta(\omega_{as})&= \frac{1}{\pi} \int_{-\infty}^{\infty}\dd{\omega_l} \tilde{\mathscr{G}}_{\beta\alpha}(\omega_{as}-\omega_{v}) \frac{\mathcal{E}_{l,\alpha} (\omega_l-\omega_{l}^{(0)}){\rm e}^{i(\omega_l-\omega_l^{(0)})\tau}}{\Gamma^2+(\omega_{as}-\omega_{v}-\omega_l)^2}\Gamma\rho_{21}(0),\nonumber\\
\mu^{(q)}_\beta(\omega_{as})&= \frac{ik_\gamma}{3\pi} \int_{-\infty}^{\infty}\dd{\omega_l}\tilde{A}_{\beta,\gamma\alpha}(\omega_{as}-\omega_{v}) \frac{\mathcal{E}_{l,\alpha} (\omega_l-\omega_{l}^{(0)}){\rm e}^{i(\omega_l-\omega_l^{(0)})\tau}}{\Gamma^2+(\omega_{as}-\omega_{v}-\omega_l)^2}\Gamma\rho_{21}(0),\nonumber\\
q_{\gamma\beta}(\omega_{as})&= \frac{1}{\pi} \int_{-\infty}^{\infty}\dd{\omega_l}\tilde{\mathscr{A}}_{\alpha,\gamma\beta}(\omega_{as}-\omega_{v}) \frac{\mathcal{E}_{l,\alpha} (\omega_l-\omega_{l}^{(0)}){\rm e}^{i(\omega_l-\omega_l^{(0)})\tau}}{\Gamma^2+(\omega_{as}-\omega_{v}-\omega_l)^2}\Gamma\rho_{21}(0),
\end{align}
where $\rho_{21}(0)$ is the off-diagonal term of the molecular density matrix at initial time $0$ and $k_\gamma$ is the $\gamma$th component of the wavevector of the incident laser pulse.
Explicit forms of the electric dipole polarizability tensor $\pmb{\tilde{\alpha}}$, electric dipole--magnetic dipole optical activity tensors $\{\pmb{\tilde{G}}$, $\pmb{\tilde{\mathscr{G}}}\}$ and electric dipole--electric quadrupole optical activity tensors $\{\pmb{\tilde{A}}$, $\pmb{\tilde{\mathscr{A}}}\}$ are given in Appendix~\ref{app_a}.
The tilde in these expressions indicates that the tensors are complex valued. The superscripts $e$, $m$ and $q$ in Eq.~\eqref{induced} represent the perturbation due to electric dipole, magnetic dipole and electric quadrupole interactions, respectively. Here, we focus on the contribution coming from the molecular coherence $\rho_{21}$ as the contributions from $\rho_{22}$ and $\rho_{11}$ will be small.
It should be noted that the conventional calculation of the ROA involves the ground state population $\rho_{11}$ and hence the tensors arising from the use of the molecular coherence are somewhat different due to the use of the initial conditions.

Unlike traditional ROA signals which are due to spontaneous Raman processes, we concentrate here on the CARS-ROA signals. This is possible as the system in prepared with significant molecular coherence and all the molecules contribute coherently to the signal. Thus the CARS-ROA signal is proportional to $N^2$, where $N$ is the number of molecules in the laser beam:
\begin{align}\label{signal}
I(\text{CARS-ROA})=N^2\vert E(\omega_{as})\vert^2,
\end{align}
where $E(\omega_{as})$ is electric field of scattered anti-Stokes light at frequency $\omega_{as}$.
The signal \eqref{signal} is to be compared with the incoherent ROA signal
\begin{align}
\frac{I(\text{CARS-ROA})}{I(\text{ROA})}\cong N\frac{\vert \rho_{21}\vert^2}{\rho_{11}},
\end{align}
and clearly CARS-ROA can be many orders larger than $I(ROA)$. Here, we have assumed that the laser pulse is applied immediately after molecular coherence has been created. A more flexible scenario would be to apply the laser pulse after a delay $\tau$. Then one needs to take into account the decay of the molecular coherence $\rho_{21}(t)\rightarrow \rho_{21}(0)\exp(-i\omega_{v} t-\Gamma t)$, where $\Gamma = 1/T_2$ is the molecular dephasing constant. For this scenario, it is more convenient to work in the frequency domain where the molecular coherence has the form
$
\left(1/\sqrt{2\pi}\right)\int_{-\infty}^\infty dt \theta(t)\rho_{21}(t){\rm e}^{i\omega t}
$.
Therefore, the anti-Stokes signal at $\omega_{as}$ will be produced from the laser pulse at frequency $\omega_{as}-\omega_v$ and the molecular coherence at $\omega_v$.

Scattered anti-Stokes field only at forward direction $z$ is under consideration. In this case, right- and left-circularly polarized components $E^x_{R/L}=\mathbf{e}_{L/R}\cdot\mathbf{E}^x$ of scattered anti-Stokes field $\mathbf{E}^x$ are found to be proportional to the molecular coherence $\rho_{21}$ between ground $\vert 1\rangle$ and excited $\vert 2\rangle$ states when we omit frequency dependence of polarizability and optical activity tensors.
Clearly, the net signal will be given by
\begin{align}\label{field1}
E^x_{R/L}(\omega_{as},\tau)\propto &
\frac{N}{\sqrt{2}}
\left(
\alpha_{xx}
\pm i\alpha_{yx}
+\frac{ik_l}{3}A_{x,zx}
-\frac{ik_{as}}{3}A_{x,zx}
\right.\nonumber\\
\pm &
\left.
\frac{1}{c}G'_{yy}\mp \frac{k_l}{3}A_{y, z x}
\pm 
\frac{1}{c}G'_{xx}
\pm 
\frac{k_{as}}{3} 
A_{x,yz}
\right)
F(\omega_{as},\tau),
\end{align}
where
\begin{align}\label{Fff}
F(\omega_{as},\tau)=
\frac{1}{\pi} \int_{-\infty}^{\infty}\dd{\omega_l} \frac{\mathcal{E}_{l,\alpha} (\omega_l-\omega_{l}^{(0)})
{\rm e}^{i(\omega_l-\omega_l^{(0)})\tau}}{\Gamma^2+(\omega_{as}-\omega_{v}-\omega_l)^2}
\Gamma\rho_{21}(0).
\end{align}
Here, the sign at the top (bottom) refers to right- (left-) circularly polarized component of the scattered field.
The tensors $\pmb{\alpha}$, $\pmb{G}'$ and $\pmb{A}$ are now real-valued (see Appendix~\ref{app_a} for details) and their values are replaced by the values at the central value $\omega_l^{(0)}$ of the incident laser pulse. 
%
%

For simplicity, finite duration of pump and Stokes fields is disregarded in Eq.~\eqref{Fff}. Its inclusion will modify $\rho_{21}(0)$ to
\begin{align}\label{rho00}
\rho_{21}(0)&\simeq \sum_{\{\vert 3\rangle\}}\frac{i\langle 2 \vert \hat{\mu}_\beta \vert 3\rangle \langle 3 \vert \hat{\mu}_\alpha \vert 1\rangle}{2\hbar^2}
\int_{-\infty}^\infty \dd{\omega_p}\frac{\mathcal{E}_{s,\beta}(\omega_{v}-\omega_p+\omega_s^{(0)})\mathcal{E}_{p,\alpha}(\omega_p-\omega_p^{(0)})}{\omega_{31}-\omega_p-i\Gamma_{3}}
\end{align}
where $\mathcal{E}_{p,\alpha}$ and $\mathcal{E}_{s,\beta}$ are the electric fields of pump and Stokes, respectively and $\Gamma_3$ is decay constant of the levels $\{\vert 3\rangle\}$.
The result Eq.~\eqref{rho00} also holds for the scheme of Fig.~\ref{fig2}b.
Detailed derivation of this expression is given in Appendix~\ref{app_e}.

\begin{figure}
\begin{center}
\includegraphics[scale=1]{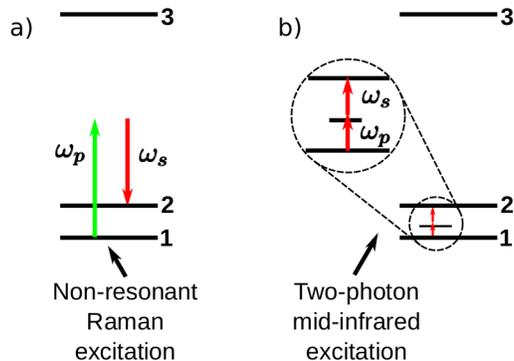}
\caption{Two methods to create the molecular coherence: a) non-resonant Raman excitation and b) two-photon mid-infrared excitation.}\label{fig2}
\end{center}
\end{figure}

We close this section by giving a brief discussion of some methods for producing molecular coherence.
One method, as shown in Fig.~\ref{fig2}a, uses laser pulses which cause off-resonant Raman transition between the levels $\vert 1\rangle$ and $\vert 2\rangle$ as say in time-resolved CARS \cite{Schmitt1998,Cheng2001,Scully2002,Volkmer2002,Hellerer2004,Prince2006}. However, much stronger pulses need to be used to produce strong molecular coherence between the levels $\vert 1\rangle$ and $\vert 2\rangle$.

Another way to produce maximum molecular coherence could be chirped-pulse adiabatic control where molecular coherence is efficiently controlled by linearly chirped pump and Stokes pulses or constant chirp in the pump and sign flipped chirp in the Stokes pulse~\cite{Malinovskaya2007}. This is a robust and efficient control on molecular coherence.   

The alternative method uses mid-infrared fields to create a two-photon transition between the levels $\vert 1\rangle$ and $\vert 2\rangle$ (see Fig.~\ref{fig2}b). The two-photon transition-created molecular coherence survives even after averaging over the orientations. This method can produce one to two orders stronger molecular coherence in comparison with spontaneous technique \cite{Traverso2018}. In addition, this method uses a low-power infrared laser, and therefore it may be considered a safe method for biological samples.

In the next sections we discuss various special cases of CARS-ROA by pre- and post-selecting polarizations.

\section{Lin-Cir and Cir-Lin CARS-ROA}
When the laser pulse is linearly polarized along the $x$-axis, the circular components of the scattered emission can be measured for studying chiral molecules. We name this configuration of measurement Lin-Cir CARS-ROA. In this case,
using expression~\eqref{field1} for the anti-Stokes electric field, the difference spectrum $I^x_R-I^x_L$ and circular intensity sum $I^x_R+I^x_L$ are found to be
\begin{align}\label{diff1}
I^x_R-I^x_L &\propto
\left(\frac{180aG'+4\gamma^2(G')}{45c}
-\frac{6\gamma^2(A)-2(\omega_{as}/\omega_l)\gamma^2(A)}{45c}
\right)
N^2\vert F\vert^2
\end{align}
and
\begin{align}\label{sum1}
I^x_R+I^x_L&\propto
\frac{45a^2+7\gamma^2(\alpha)}{45}N^2\vert F\vert^2.
\end{align}
The real-valued tensor invariants $a$, $G'$, $\gamma^2(\alpha)$, $\gamma^2(G')$ and $\gamma^2(A)$ are defined by \cite{Long2002,Nafie2011}
\begin{align}
a^2 &=
\frac{1}{9}\alpha_{\lambda_1\lambda_1}\alpha_{\lambda_2\lambda_2},\nonumber\\
\gamma^2(\alpha) &= \frac{1}{2}(3\alpha_{\lambda_1\lambda_2}\alpha_{\lambda_1\lambda_2}-\alpha_{\lambda_1\lambda_1}\alpha_{\lambda_2\lambda_2}),\nonumber\\
aG' &= \frac{1}{9}\alpha_{\lambda_1\lambda_1}G'_{\lambda_2\lambda_2},\nonumber\\
\gamma^2(G') &= \frac{1}{2}(3\alpha_{\lambda_1\lambda_2}G'_{\lambda_1\lambda_2}-\alpha_{\lambda_1\lambda_1}G'_{\lambda_2\lambda_2}),\nonumber\\
\gamma^2(A) &= \frac{\omega_l}{2} \epsilon_{\lambda_2\lambda_3\lambda_4}\alpha_{\lambda_1\lambda_2}A_{\lambda_3,\lambda_4\lambda_1}.
\end{align}
Equations~\eqref{diff1} and \eqref{sum1} are very similar to the signal terms found by L.~D.~Barron \cite{Barron1971}. The small difference disappears upon making the nonphysical assumption $\omega_l=\omega_{as}$. However, these equations now have more deep physical meaning since these signals depend on molecular coherence $\rho_{21}$ and most importantly on $N^2$. In addition, note that the difference spectrum~\eqref{diff1} and circular intensity sum~\eqref{sum1} are $\tau$-dependent.

Using Eqs.~\eqref{diff1} and \eqref{sum1} along with the fact that the generated molecular coherence for schemes of Fig.~\ref{fig2} can be two orders higher than that of spontaneous Raman (see for example Refs.~\cite{Scully2002,Traverso2018}), we reach the conclusion that $I^x_R\mp I^x_L$ for Lin-Cir CARS-ROA are at most $10^4$ times larger than that for ROA. Here, we should note that not only the difference spectrum is enhanced but also the circular intensity sum. Therefore, the circular intensity difference $\Delta = (I^x_R - I^x_L)/(I^x_R + I^x_L)$ does not change.
However, the enhancement in absolute magnitude of difference spectrum allows us to significantly increase the signal-to-noise ratio for the chiral signal.

For circularly polarized laser and linearly polarized scattered light, we call this configuration Cir-Lin CARS-ROA. The difference spectrum can be found as
\begin{align}\label{arvanneg}
I^R_x-I^L_x &\propto
\left(\frac{180aG'+4\gamma^2(G')}{45c}
-\frac{6(\omega_{as}/\omega_l)\gamma^2(A)-2\gamma^2(A)}{45c}
\right)
N^2\vert F\vert^2,
\end{align}
and the circular intensity sum $I^R_x+I^L_x$ is the same as
that of Lin-Cir CARS-ROA (see Appendix~\ref{app_b} and \ref{app_c} for details).
Since the dependence on molecular coherence remains in Eq.~\eqref{arvanneg}, the enhancement factors are the same for Lin-Cir and Cir-Lin CARS-ROA.

\section{Heterodyne measurements with a local oscillator}
In the case of heterodyne measurement with a local oscillator $\mathbf{E}_{LO}$ of frequency $\omega_{as}$, the heterodyne signal is defined as
\begin{align}
I^x_{R/L}\propto\vert E^x_{R/L}+E_{LO}\vert^2,
\end{align}
where $E_{LO}=E_{LO,R/L}=\mathbf{e}_{L/R}\cdot \mathbf{E}_{LO}$, that is, we assume right- and left-circularly polarized components of the local oscillator are equal to each other.
The local oscillator can be obtained from the laser pulse used to create CARS-ROA by modulating it.
Then, the difference spectrum and circular intensity sum are found to be
\begin{align}\label{diff}
I^x_R-I^x_L &\propto
\left(\frac{180aG'+4\gamma^2(G')}{45c}
-\frac{6\gamma^2(A)-2(\omega_{as}/\omega_l)\gamma^2(A)}{45c}
\right)
N^2\vert F\vert^2
\nonumber\\
&+
\frac{8}{\sqrt{2}c}G' N \operatorname{Re}(F E_{LO}^*)
\end{align}
and
\begin{align}\label{sum}
I^x_R+I^x_L&\propto
\frac{45a^2+7\gamma^2(\alpha)}{45}N^2\vert F\vert^2
+
\frac{4}{\sqrt{2}} a N \operatorname{Re}(F E_{LO}^*).
\end{align}
The first terms in Eqs.~\eqref{diff} and \eqref{sum} are Lin-Cir CARS-ROA signals whereas the second terms are heterodyne signals.
In order to recover the heterodyne signal we need to remove the first terms in Eqs.~\eqref{diff} and \eqref{sum}.
The way to do so is to measure signal Eqs.~\eqref{diff} and \eqref{sum} with two different phases of $E_{LO}$, namely $\phi$ and $\phi+\pi$ phases, and to subtract the two heterodyne difference spectra (similar method for CARS was reported in Ref.~\cite{Lu2008}). This enables us to measure only the heterodyne chiral parameter $G'$ and achiral parameter $a$, that are,
\begin{align}\label{doubled}
(I_R^x-&I_L^x)_{\phi}-
(I_R^x-I_L^x)_{\phi+\pi}
\propto
\frac{16}{\sqrt{2}c}G'N \operatorname{Re}(F \vert E_{LO}\vert \exp (i\phi)),
\nonumber\\
(I_R^x+&I_L^x)_{\phi}-
(I_R^x+I_L^x)_{\phi+\pi}
\propto
\frac{8}{\sqrt{2}} aN \operatorname{Re}(F \vert E_{LO}\vert \exp (i\phi)).
\end{align}
Note the striking aspect of Eq.~\eqref{doubled}: it directly determines the ratio of tensor invariants $G'$ and $a$. This would be the first direct measurement of this ratio. Furthermore, the expressions in Eq.~\eqref{doubled} still depend on the molecular coherence $\rho_{21}$ via $F$. Hence, the heterodyne chiral signal \eqref{doubled} can be enhanced by factor of $10^2$ at most. An estimate of the magnetic and quadrupole contributions is given in the experiments of D.~Che and L.~A.~Nafie \cite{Che1992}, where it is reported that such contributions to ROA signals are about 1000 times smaller than the electric dipole one.

\section{Discussion and conclusions}
We predict that, depending on the type of measurement configuration, a chiral signal $10^2-10^4$ times stronger in magnitude can be obtained by creating strong molecular coherence. This is for Lin-Cir and Cir-Lin CARS-ROA.
As shown above, molecular coherence enhances not only the difference spectrum but also the intensity sum too. This is one of the key points of the present paper.

Furthermore, we present a new heterodyne measurement scheme that allows us to experimentally determine the ratio of tensor invariants $G'$ and $a$. A general expression for the CARS-ROA signal consists of not only the actual heterodyne signal but also the signal coming from interference between chiral and achiral terms which includes the product of $aG'$ and other anisotropic tensor invariants $\gamma^2(G')$ and $\gamma^2(A)$.
This interference term makes the heterodyne measurement less accurate. To overcome this difficulty one may measure the heterodyne difference spectrum and circular intensity sum with two different phases of local oscillator and eliminate the interference term coming from chiral and achiral terms by subtracting the measured signals. As a result, the measured signal only depends on $G'$ and $a$, which enables us to obtain the ratio between $G'$ and $a$. 


For materials that are not heat resistant, such as biological molecules, we suggest either the two-photon mid-infrared excitation method or resonant Raman excitation method for inducing molecular coherence. Both methods use low power of excitation, and consequently, they are expected to be safe for most samples.

In summary, we demonstrated that the chiral signal could be enhanced via molecular coherence in comparison with ROA signals where molecular coherence essentially plays no role. In the case of non-resonant Raman excitation, the enhancement factor is estimated up to $10^4$. Such benefit of the molecular coherence provides a new alternative technique for investigating chiral molecules in stereochemistry and biochemistry.
Although we have restricted to the lowest-order optically active processes, we expect molecular coherence to play an equally important role in higher-order optically active processes.

\section*{Acknowledgments}
GSA thanks L.~D.~Barron for valuable comments.
We are grateful to the Air Force Office of Scientific Research (Award No. FA9550-18-1-0141) and the Office of Naval Research (Award No. N00014-16-1-3054). T. B. is supported by the Herman F. Heep and Minnie Belle Heep Texas A\&M University Endowed Fund held/administered by the Texas A\&M Foundation.

\appendix
\section*{Appendix}
\section{Initial molecular coherence}\label{app_e}
\begin{figure}
\begin{center}
\includegraphics[scale=1]{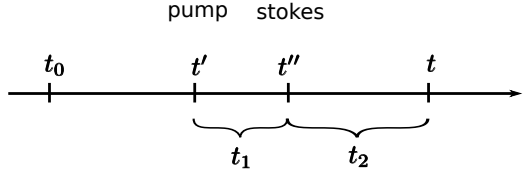}
\caption{New time variables $t_1$ and $t_2$ used in Eq.~\eqref{rhonew}}\label{fig3}
\end{center}
\end{figure}

For completeness and the convenience of the reader, details of derivation of Eq.~\eqref{rho00} are presented in this appendix. Problem is to find time dependent molecular coherence $\rho_{21}(t)$ created by pump $\mathbf{E}_{p}(t)$ and Stokes $\mathbf{E}_s(t)$ pulses. Using semiclassical light matter Hamiltonian $H^{(I)}(t)$ in the interaction picture we obtain molecular density operator $\hat{\rho}^{(I)}(t)$ and its matrix element $\rho^{(I)}_{21}(t)$ in the interaction picture as follow 
\begin{align}
\hat{\rho}^{(I)}(t)&\simeq \left(-\frac{i}{\hbar}\right)^2
\int_{t_0}^t \dd{t''}\int_{t_0}^{t''}\dd{t'}[\hat{H}^{(I)}(t''),[\hat{H}^{(I)}(t'),\hat{\rho}^{(I)}(t')]]
\end{align}
and
\begin{align}\label{rho125}
\rho^{(I)}_{21}(t)&\simeq \left(-\frac{i}{\hbar}\right)^2 \langle 2 \vert \hat{\mu}_\beta \vert 3\rangle \langle 3 \vert \hat{\mu}_\alpha \vert 1\rangle
\int_{t_0}^t \dd{t''}\int_{t_0}^{t''}\dd{t'}
{\rm e}^{i\omega_{23}t''}{\rm e}^{i\omega_{31}t'}
E_{s,\beta}(t'')E_{p,\alpha}(t').
\end{align}
Here, initial condition for density matrix is $\rho_{11}(t_0)=1$.
New variables $t_1$ and $t_2$ defined in Fig.~\ref{fig3} can simplify double integral in Eq.~\eqref{rho125} to
\begin{align}\label{rhonew}
\rho_{21}(t)\simeq &\left(-\frac{i}{\hbar}\right)^2 
\langle 2 \vert \hat{\mu}_\beta \vert 3\rangle \langle 3 \vert \hat{\mu}_\alpha \vert 1\rangle\nonumber\\
\times &\int_{-\infty}^{\infty}\dd{t_2}\theta(t_2){\rm e}^{-i\omega_{21}t_2}E_{s,\beta}(t-t_2)
\int_{-\infty}^{\infty}\dd{t_1}\theta(t_1){\rm e}^{-i\omega_{31}t_1}E_{p,\alpha}(t-t_1-t_2),
\end{align}
where $\theta(t)$ is Heaviside step function. With the standard definition of Fourier transform of a function $F(t)$ as
\begin{align}
F(t)=\frac{1}{\sqrt{2\pi}}\int_{-\infty}^{\infty}\dd{\omega}F(\omega){\rm e}^{-i\omega t},\nonumber\\
F(\omega)=\frac{1}{\sqrt{2\pi}}\int_{-\infty}^{\infty}\dd{t}F(t){\rm e}^{i\omega t}
\end{align}
and using convolution theorem, finally, we obtain dynamics of molecular coherence as follows
\begin{align}\label{rhodyn}
\rho_{21}(t)&\simeq \frac{\langle 2 \vert \hat{\mu}_\beta \vert 3\rangle \langle 3 \vert \hat{\mu}_\alpha \vert 1\rangle}{2\pi\hbar^2}
\int_{-\infty}^\infty \dd{\omega_p}\int_{-\infty}^\infty \dd{\Omega}
\frac{{\rm e}^{-i\Omega t}E_{s,\beta}(\Omega-\omega_p)E_{p,\alpha}(\omega_p)}{(\omega_p-\omega_{31}+i\Gamma_{3})(\Omega-\omega_{21}+i\Gamma)}.
\end{align}
When pump and Stokes pulses are nearly in two-photon resonance with transition $\omega_{21}$ the factor $1/(\Omega-\omega_{21}+i\Gamma)$ in Eq.~\eqref{rhodyn} yields $-i\pi\delta(\Omega-\omega_{21})$, and consequently, Eq.~\eqref{rhodyn} with summation over all possible excited states $\{\vert 3\rangle\}$ provides us Eq.~\eqref{rho00}. 

\section{Polarizability and optical activity tensors}\label{app_a}
Our starting point is the field-matter interaction Hamiltonian \eqref{Hamiltonian0}.
Once we calculate the polarizability tensor with electric dipole interactions, the procedure for derivation of optical activity tensors is straightforward.

The induced polarization of a single molecule is given by
\begin{align}\label{indpol}
\pmb{\mu}(t)&=\operatorname{Tr}[\hat{\pmb{\mu}}\hat{\rho}(t)]
=\sum_{\{3\}}
\left(\mu_{13}\rho_{31}(t)+\mu_{23}\rho_{32}(t)
+\mu_{31}\rho_{13}(t)+\mu_{32}\rho_{23}(t)\right),
\end{align}
where $\hat{\rho}(t)$ is a density matrix of a single molecule. The time evolution of $\hat{\rho}(t)$ is governed by the von Neumann equation $d\hat{\rho}(t)/dt=(-i/\hbar)[\hat{H}(t),\hat{\rho}(t)]$, and exploiting first-order perturbation theory, the density matrix elements $\rho_{31}(t)$ and $\rho_{32}(t)$ at time $t$ are found to be
\begin{align}\label{per}
\rho_{31}(t)&={\rm e}^{-i\omega_{31}t}\left(-\frac{i}{\hbar}\right)\int_{0}^t \dd{t'} H^{(I)}_{32}(t')\rho_{21}(0),\nonumber\\
\rho_{23}(t)&={\rm e}^{-i\omega_{23}t}\left(\frac{i}{\hbar}\right)\int_{0}^t \dd{t'} \rho_{21}(0) H^{(I)}_{13}(t').
\end{align}
Here, we keep only anti-Stokes Raman terms. Matrix elements of the Hamiltonian in the interaction picture are denoted by $H^{(I)}_{32}(t')$ and $H^{(I)}_{31}(t')$ and their explicit form is given by
\begin{align}
H^{(I)}_{32}(t')=-\langle 3\vert \hat{\mu}_\alpha\vert 2\rangle E_{l,\alpha}(t'){\rm e}^{-\Gamma \vert t'\vert}{\rm e}^{i\omega_{32} t'},\\
H^{(I)}_{13}(t')=-\langle 1\vert \hat{\mu}_\alpha\vert 3\rangle E_{l,\alpha}(t'){\rm e}^{-\Gamma \vert t'\vert}{\rm e}^{i\omega_{13} t'},
\end{align}
where dephasing constant $\Gamma$ is introduced as a part of electric field as $E_{l,\alpha}(t')\exp(-\Gamma \vert t'\vert)$.
Then spectral decomposition of electric field is given by
\begin{align}
E_{l,\alpha}(t'){\rm e}^{-\Gamma \vert t'\vert}=\frac{1}{\sqrt{2\pi}}
\int_{-\infty}^\infty \dd{\omega_l}
\left[
\frac{1}{\sqrt{2\pi}}\int_{-\infty}^\infty\dd{\omega'}E_{l,\alpha}(\omega')\frac{\sqrt{2/\pi}\Gamma}{\Gamma^2+(\omega_l-\omega')^2}
\right]
{\rm e}^{-i\omega_l t'}.
\end{align}
%
Furthermore, Eqs.~\eqref{per} yield
\begin{align}
\rho_{31}(t)&=\frac{1}{2\pi\hbar}\int_{-\infty}^\infty \dd{\omega_l}
\frac{\langle 3\vert \hat{\mu}_\alpha\vert 2\rangle }{\omega_{32}-\omega_l-i\Gamma_3}
\int_{-\infty}^\infty\dd{\omega'}E_{l,\alpha}(\omega')\frac{\sqrt{2/\pi}\Gamma}{\Gamma^2+(\omega_l-\omega')^2}
\rho_{21}(0){\rm e}^{-i(\omega_{21}+\omega_l)t}
,\nonumber\\
\rho_{23}(t)&=\frac{1}{2\pi\hbar}\int_{-\infty}^\infty \dd{\omega_l}
\frac{\langle 1\vert \hat{\mu}_\alpha\vert 3\rangle }{\omega_{31}+\omega_{l}+i\Gamma_3}
\int_{-\infty}^\infty\dd{\omega'}E_{l,\alpha}(\omega')\frac{\sqrt{2/\pi}\Gamma}{\Gamma^2+(\omega_l-\omega')^2}
\rho_{21}(0){\rm e}^{-i(\omega_{21}+\omega_l)t}
,
\end{align}
Plugging $\rho_{31}(t)$ and $\rho_{23}(t)$ into Eq.~\eqref{indpol}, we obtain $\beta$th component of induced electric dipole moment due to electric dipole perturbation as follows:
\begin{align}\label{mubt}
\mu^{(e)}_\beta(t)=\frac{1}{2\pi}\int_{-\infty}^{\infty}\dd{\omega_l}
\tilde{\alpha}_{\beta\alpha}(\omega_l)
\int_{-\infty}^\infty\dd{\omega'}E_{l,\alpha}(\omega')\frac{\sqrt{2/\pi}\Gamma}{\Gamma^2+(\omega_l-\omega')^2}
 {\rm e}^{-i\omega_l t}\rho_{21}(0){\rm e}^{-i \omega_{21} t}.
\end{align}
Here, superscript $(e)$ indicates that the electric dipole moment is induced by electric dipole perturbation. 
The polarizability tensor is defined by
\begin{align}\label{alp1}
\tilde{\alpha}_{\beta\alpha}(\omega_l)=
\frac{1}{\hbar}
\sum_{\{3\}}
\left(
\frac{\langle 1\vert \hat{\mu}_\beta \vert 3\rangle\langle 3\vert \hat{\mu}_\alpha \vert 2\rangle}{\omega_{32}-\omega_{l}-i\Gamma_3}
+
\frac{\langle 1\vert \hat{\mu}_\alpha \vert 3\rangle\langle 3\vert \hat{\mu}_\beta \vert 2\rangle}{\omega_{31}+\omega_{l}+i\Gamma_3}
\right),
\end{align}
where summation over all possible excited states $\{\vert 3\rangle\}$ is taken.
In the same way, $\beta$th component of induced electric dipole moments $\mu^{(m)}_\beta(t)$ due to magnetic dipole perturbation, $\mu^{(q)}_\beta(t)$ due to electric quadrupole perturbation, induced magnetic dipole $m_\beta(t)$ and electric quadrupole $q_{\gamma\beta}(t)$ moments due to electric dipole perturbation are, respectively
\begin{align}\label{induced2}
\mu_\beta^{(m)}(t) &= \frac{1}{2\pi}\int_{-\infty}^\infty \dd{\omega_l}\tilde{G}_{\beta\alpha}(\omega_l)
\int_{-\infty}^\infty\dd{\omega'}B_{l,\alpha}(\omega')\frac{\sqrt{2/\pi}\Gamma}{\Gamma^2+(\omega_l-\omega')^2}
{\rm e}^{-i\omega_l t}\rho_{21}(0){\rm e}^{-i\omega_{v}t},\nonumber\\
m_\beta(t) &= \frac{1}{2\pi}\int_{-\infty}^\infty \dd{\omega_l}\tilde{\mathscr{G}}_{\beta\alpha}(\omega_l)
\int_{-\infty}^\infty\dd{\omega'}E_{l,\alpha}(\omega')\frac{\sqrt{2/\pi}\Gamma}{\Gamma^2+(\omega_l-\omega')^2}
{\rm e}^{-i\omega_l t}\rho_{21}(0){\rm e}^{-i\omega_{v}t},\nonumber\\
\mu_\beta^{(q)}(t) &= \frac{1}{2\pi}\int_{-\infty}^\infty \dd{\omega_l}\frac{i}{3}\tilde{A}_{\beta,\gamma\alpha}(\omega_l)k_\gamma \int_{-\infty}^\infty\dd{\omega'}E_{l,\alpha}(\omega')\frac{\sqrt{2/\pi}\Gamma}{\Gamma^2+(\omega_l-\omega')^2}
{\rm e}^{-i\omega_l t}\rho_{21}(0){\rm e}^{-i\omega_{v}t},\nonumber\\
q_{\gamma\beta}(t) &= \frac{1}{2\pi}\int_{-\infty}^\infty \dd{\omega_l}\tilde{\mathscr{A}}_{\alpha,\gamma\beta}(\omega_l)
\int_{-\infty}^\infty\dd{\omega'}E_{l,\alpha}(\omega')\frac{\sqrt{2/\pi}\Gamma}{\Gamma^2+(\omega_l-\omega')^2}
{\rm e}^{-i\omega_l t}\rho_{21}(0){\rm e}^{-i\omega_{v}t},
\end{align}
where superscripts $m$ and $q$ denote magnetic dipole and electric quadrupole perturbations, respectively, and $k_\gamma$ is the $\gamma$th component of wave-vector of the incident laser pulse.
In Eq.~\eqref{induced2}, the electric dipole-magnetic dipole optical activity tensors $\{\pmb{\tilde{G}}$, $\pmb{\tilde{\mathscr{G}}}\}$ and electric dipole-electric quadrupole optical activity tensors $\{\pmb{\tilde{A}}$, $\pmb{\tilde{\mathscr{A}}}\}$ are defined as follows:
\begin{align}\label{alp2}
\tilde{G}_{\beta\alpha}=
\frac{1}{\hbar}
\sum_{\{3\}}
\left(
\frac{\langle 1\vert \hat{\mu}_\beta \vert 3\rangle\langle 3\vert \hat{m}_\alpha \vert 2\rangle}{\omega_{32}-\omega_{l}-i\Gamma_3}
+
\frac{\langle 1\vert \hat{m}_\alpha \vert 3\rangle\langle 3\vert \hat{\mu}_\beta \vert 2\rangle}{\omega_{31}+\omega_{l}+i\Gamma_3}
\right),
\end{align}
\begin{align}\label{alp3}
\tilde{\mathscr{G}}_{\beta\alpha}=
\frac{1}{\hbar}
\sum_{\{3\}}
\left(
\frac{\langle 1\vert \hat{m}_\beta \vert 3\rangle\langle 3\vert \hat{\mu}_\alpha \vert 2\rangle}{\omega_{32}-\omega_{l}-i\Gamma_3}
+
\frac{\langle 1\vert \hat{\mu}_\alpha \vert 3\rangle\langle 3\vert \hat{m}_\beta \vert 2\rangle}{\omega_{31}+\omega_{l}+i\Gamma_3}
\right),
\end{align}
\begin{align}\label{alp4}
\tilde{A}_{\beta,\gamma\alpha}=
\frac{1}{\hbar}
\sum_{\{3\}}
\left(
\frac{\langle 1\vert \hat{\mu}_\beta\vert 3\rangle\langle 3\vert \hat{q}_{\gamma\alpha}\vert 2\rangle}{\omega_{32}-\omega_{l}-i\Gamma_3}
+
\frac{\langle 1\vert \hat{q}_{\gamma\alpha}\vert 3\rangle\langle 3\vert \hat{\mu}_\beta\vert 2 \rangle}{\omega_{31}+\omega_{l}+i
\Gamma_3}
\right),
\end{align}
\begin{align}\label{alp5}
\tilde{\mathscr{A}}_{\alpha,\gamma\beta}=
\frac{1}{\hbar}
\sum_{\{3\}}
\left(
\frac{\langle 1\vert \hat{q}_{\gamma\beta} \vert 3\rangle\langle 3\vert \hat{\mu}_\alpha \vert 2\rangle}{\omega_{32}-\omega_{l}-i\Gamma_3}
+
\frac{\langle 1\vert \hat{\mu}_\alpha \vert 3\rangle\langle 3\vert \hat{q}_{\gamma\beta} \vert 2\rangle}{\omega_{31}+\omega_{l}+i\Gamma_3}
\right).
\end{align}
In the frequency domain, induced dipole moment due to electric dipole perturbation is determined by Fourier transform of Eq.~\eqref{mubt} as follows:
\begin{align}\label{moment1}
\mu^{(e)}_\beta(\omega_{as})= \frac{1}{\pi} \int_{-\infty}^{\infty}\dd{\omega_l} \tilde{\alpha}_{\beta\alpha}(\omega_{as}-\omega_{v}) \frac{\mathcal{E}_{l,\alpha} (\omega_l-\omega_{l}^{(0)})
{\rm e}^{i(\omega_l-\omega_l^{(0)})\tau}}{\Gamma^2+(\omega_{as}-\omega_{v}-\omega_l)^2}
\Gamma\rho_{21}(0),
\end{align}
where vibrational frequency $\omega_v=\omega_{21}$ and the electric field of the laser pulse in time domain is defined as $E_{l,\alpha}(t)=\mathcal{E}_{l,\alpha}(t-\tau){\rm e}^{-i\omega_l^{(0)} t}$ and its Fourier transform is given by $E_{l,\alpha}(\omega_l)=\mathcal{E}_{l,\alpha} (\omega_l-\omega_l^{(0)}){\rm e}^{i(\omega_l-\omega_l^{(0)}) \tau}$.
Similarly, other induced dipole and quadrupole moments are written in the frequency domain as follows:
\begin{align}\label{moment2}
\mu^{(m)}_\beta(\omega_{as})&= \frac{1}{\pi} \int_{-\infty}^{\infty}\dd{\omega_l}\tilde{G}_{\beta\alpha}(\omega_{as}-\omega_{v}) \frac{\mathcal{B}_{l,\alpha} (\omega_l-\omega_{l}^{(0)}){\rm e}^{i(\omega_l-\omega_l^{(0)})\tau}}{\Gamma^2+(\omega_{as}-\omega_{v}-\omega_l)^2}\Gamma\rho_{21}(0),\nonumber\\
m_\beta(\omega_{as})&= \frac{1}{\pi} \int_{-\infty}^{\infty}\dd{\omega_l} \tilde{\mathscr{G}}_{\beta\alpha}(\omega_{as}-\omega_{v}) \frac{\mathcal{E}_{l,\alpha} (\omega_l-\omega_{l}^{(0)}){\rm e}^{i(\omega_l-\omega_l^{(0)})\tau}}{\Gamma^2+(\omega_{as}-\omega_{v}-\omega_l)^2}\Gamma\rho_{21}(0),\nonumber\\
\mu^{(q)}_\beta(\omega_{as})&= \frac{ik_\gamma}{3\pi} \int_{-\infty}^{\infty}\dd{\omega_l}\tilde{A}_{\beta,\gamma\alpha}(\omega_{as}-\omega_{v}) \frac{\mathcal{E}_{l,\alpha} (\omega_l-\omega_{l}^{(0)}){\rm e}^{i(\omega_l-\omega_l^{(0)})\tau}}{\Gamma^2+(\omega_{as}-\omega_{v}-\omega_l)^2}\Gamma\rho_{21}(0),\nonumber\\
q_{\gamma\beta}(\omega_{as})&= \frac{1}{\pi} \int_{-\infty}^{\infty}\dd{\omega_l}\tilde{\mathscr{A}}_{\alpha,\gamma\beta}(\omega_{as}-\omega_{v}) \frac{\mathcal{E}_{l,\alpha} (\omega_l-\omega_{l}^{(0)}){\rm e}^{i(\omega_l-\omega_l^{(0)})\tau}}{\Gamma^2+(\omega_{as}-\omega_{v}-\omega_l)^2}\Gamma\rho_{21}(0),
\end{align}
where $\mathcal{B}_{l,\alpha}(\omega-\omega_l^{(0)})$ is a Fourier transform of envelope function of magnetic field.
Here, $\vert 1\rangle$, $\vert 2\rangle$ and $\vert 3\rangle$ are ground and excited electro-vibrational states of a molecule. 
Generally, the tensors \eqref{alp1} and \eqref{alp2}--\eqref{alp5} are complex valued (tilde of these tensors means that they are complex), but commonly used assumptions are: (a)
the Born--Oppenheimer approximation, where each molecular state is a direct product of electronic and vibrational wavefunctions; and (b)
we can always choose real wavefunctions in the absence of external magnetic field, so allowing us to assume the electric dipole and quadrupole moment operators are purely real and 
the magnetic dipole moment operator is purely imaginary.
With these two assumptions, we have only three tensors (without tilde), namely real-valued tensors $(\alpha_{\beta\alpha})_{12}$ and $(A_{\beta,\gamma\alpha})_{12}=(\mathscr{A}_{\beta,\gamma\alpha})_{12}$, and imaginary-valued tensor $(G_{\beta\alpha})_{12}=(\mathscr{G}_{\alpha\beta}^*)_{12}$. Moreover, we exclude the imaginary unit $i$ from $(G_{\beta\alpha})_{12}$ by introducing the purely real tensor $(G_{\beta\alpha})'_{12}=i(G_{\beta\alpha})_{12}$.
 
The scattered electric fields $\mathbf{E}_\mu$, $\mathbf{E}_m$ and $\mathbf{E}_q$ at point $r\mathbf{n}$ in radiation zone due to electric dipole, magnetic dipole and electric quadrupole moments, respectively, are given by \cite{Jackson1975}
\begin{align}\label{emitted}
\mathbf{E}_\mu &=Z_0 \frac{c k_{as}^2}{4\pi} \frac{{\rm e}^{ik_{as}r}}{r}(\mathbf{n}\times\pmb{\mu})\times \mathbf{n},\nonumber\\
\mathbf{E}_m&=Z_0 \frac{k_{as}^2}{4\pi} \frac{{\rm e}^{ik_{as}r}}{r}(\mathbf{m}\times\mathbf{n}),\nonumber\\
\mathbf{E}_q&=Z_0 \frac{i c k_{as}^3}{12\pi} \frac{{\rm e}^{ik_{as}r}}{r} (\mathbf{q}\times\mathbf{n})\times\mathbf{n},
\end{align}
where $Z_0=\sqrt{\mu_0/\epsilon_0}$, $(\mathbf{q})_\alpha=q_{\alpha\beta}n_\beta$ and $\mathbf{n}$ is direction of emission. The wave-number for oscillating dipole and quadrupole moments is denoted by $k_{as}$.
Next, we calculate the electric fields of scattered emission for several different pulse configurations using Eq.~\eqref{emitted} along with the induced moments Eqs.~\eqref{moment1} and \eqref{moment2}.


\section{Lin-Cir CARS-ROA}\label{app_b}
Lin-Cir CARS-ROA refers to difference between right- and left-circularly polarized components of the scattered pulse when the incident probe pulse is linearly polarized. Let us choose $x$-polarized incident laser pulse $E_{l,x}(t)$ propagating along $z$ direction; then after squaring the Eq.~\eqref{field1} and averaging over random orientations of the molecules, we obtain intensities of circularly polarized components of scattered anti-Stokes emission as follows:
\begin{align}
I^x_{R/L}&\propto(E^x_{R/L}+E_{LO})(E^x_{R/L}+E_{LO})^*
\propto
\frac{45a^2+7\gamma^2(\alpha)}{90}N^2\vert F\vert^2
\pm
\frac{180aG'+4\gamma^2(G')}{90c}N^2\vert F\vert^2\nonumber\\
&\mp
\frac{6\gamma^2(A)}{90c}N^2\vert F\vert^2
\pm
\frac{\omega_{as}}{\omega_l}\frac{2\gamma^2(A)}{90c}N^2\vert F\vert^2
+
 \sqrt{2}a N \operatorname{Re}(F E_{LO}^*)
\pm
\frac{4}{\sqrt{2}c}G'N \operatorname{Re}(F E_{LO}^*),
\end{align}
where $E_{LO}=E_{LO,R/L}=(1/\sqrt{2})(\mathbf{e}_x \pm i\mathbf{e}_y)\cdot \mathbf{E}_{LO}$ is circular components of local oscillator at frequency $\omega_{as}$.

\section{Cir-Lin CARS-ROA}\label{app_c}
Cir-Lin CARS-ROA refers to measurement of linearly polarized component of scattered emission provided right- and left-circularly polarized incident pulses are present.
Thus, we consider a circularly polarized incident laser pulse propagating along $z$ direction whose field components are
\begin{align}
E_{l,x}(t)=\frac{E_{l}(t)}{\sqrt{2}},\quad
E_{l,y}(t)=\mp i \frac{E_{l}(t)}{\sqrt{2}},\quad
B_{l,x}(t)=\pm i\frac{E_{l}(t)}{\sqrt{2}c},\quad
B_{l,y}(t)= \frac{E_{l}(t)}{\sqrt{2}c},
\end{align}
where sign at the top indicates right-circularly polarized incident light whereas sign in the bottom indicates left-circularly polarized incident light.
In the same manner as Lin-Cir CARS-ROA, we obtain $x$ and $y$ components of scattered emission at forward direction $\mathbf{n}=\mathbf{e}_z$ as follows:
\begin{align}
E_{x}^{R/L}&\propto \frac{N}{\sqrt{2}}\left(\alpha_{xx} \mp i\alpha_{xy}
+\frac{ik_l}{3}A_{x,zx}-\frac{ik_{as}}{3}A_{x,zx}
\pm \frac{1}{c}G'_{xx}\pm\frac{k_l}{3}A_{x, z y}
\pm\frac{1}{c}G'_{yy}\mp\frac{k_{as}}{3}A_{y,xz}
\right)F,\nonumber\\
E_{y}^{R/L}&\propto \frac{N}{\sqrt{2}}\left(\alpha_{yx} \mp i\alpha_{yy}
\pm\frac{k_l}{3}A_{y,zy}\mp\frac{k_{as}}{3}A_{y,yz}
-\frac{i}{c}G'_{yy}+\frac{ik_l}{3}A_{y, z x}
-\frac{i}{c}G'_{xx}
-\frac{ik_{as}}{3}A_{x,yz}
\right)F.
\end{align}
After averaging over random molecular orientations, the intensity is found to be
\begin{align}
I_x^{R/L} &\propto
\left(
\frac{45a^2+7\gamma^2(\alpha)}{90}
\pm \frac{180aG'+4\gamma^2(G')}{90c}
\mp\frac{\omega_{as}}{\omega_l}\frac{6\gamma^2(A)}{90c}
\pm
\frac{2\gamma^2(A)}{90c}
\right)N^2\vert F\vert^2
\nonumber\\
&
+\sqrt{2} a N \operatorname{Re}(F E_{LO}^*) \pm \frac{4}{\sqrt{2}c}G' N \operatorname{Re}(F E_{LO}^*),
\end{align}
where $E_{LO}=\mathbf{e}_{x}\cdot \mathbf{E}_{LO}$.


\bibliography{apssamp.bib}

\end{document}